\documentclass[12pt]{revtex4-2}

\usepackage{graphicx}

\begin{document}

\title{Optically induced sieve effect for nanoparticles near a nanofiber taper}

\author{Mark Sadgrove$^{1,*}$,Takaaki Yoshino$^{2}$,Masakazu Sugawara$^{2}$, Yasuyoshi Mitsumori$^{2}$ and Keiichi Edamatsu$^{2}$}

\affiliation{$^1$Department of Physics, Faculty of Science, Tokyo University of Science, 1-3 Kagurazaka, Shinjuku-ku, Tokyo 162-8601, Japan\\
$^2$ Research Institute of Electrical Communication, Tohoku University, Sendai 980-8577, Japan}
\email{$^*$ mark.sadgrove@rs.tus.ac.jp}

\begin{abstract}
We demonstrate size selective optical trapping and transport for nanoparticles near an optical nanofiber taper. Using a two-wavelength, counter-propagating mode configuration, we show that 100 nm diameter and 150 nm diameter gold nanospheres (GNSs) are trapped by the evanescent field in the taper region at different optical powers. Conversely, when one nanoparticle species is trapped the other may be transported, leading to a sieve-like effect. Our results show that sophisticated optical manipulation can be achieved in a passive configuration by taking advantage of mode behavior in nanophotonics devices.
\end{abstract}
\maketitle
\section{Introduction}
Filtering particles by size is a requirement across a range of fields including 
fabrication of artificial quantum emitters~\cite{wang2005general,ryu2014facile,wei2014size,morita2008facile}, 
detection and separation of biological particles~\cite{van2006strategy,yeo2015microfluidic,lin2018dna,xuan2013size,svoboda1994biological},  
and gas separation~\cite{carta2013efficient}. Although mechanical techniques exist for many applications (with some remarkable recent examples
~\cite{carta2013efficient,stogin2018free}), in the nanoparticle regime, much recent
effort has been focused on optical methods~\cite{nan2018sorting,righini2007parallel,Ploschner}, due to their inherent flexibility.
One aspect of optical manipulation showing promise in this area is the use of counterpropagating optical fields, which 
allows sophisticated particle control including optical pulling type effects and the use of optical 
nonlinearity to achieve size sensitive optically induced forces~\cite{PhysRevLett.109.087402, Wada01}.

On the other hand, a separate research avenue for the manipulation of nano-size particles of many types makes use of nanostructures including nanowaveguides~\cite{kawata1996optically,southampton,skelton2012evanescent,maimaiti2015higher,Lipo} and plasmonic nanostructures~\cite{tanaka2011optical,min2013focused}. 
In particular, a very recent combination of nanowaveguide techniques with the above-mentioned counter-propagating field method,  has led to remarkable selective control of nanodiamonds near to an optical nanofiber conditional on the number of NV centers they contain~\cite{Sasaki2}. This suggests that combining multiple wavelength techniques with nanooptics may prove to be fertile ground for the realization of novel selective manipulation techniques.

Here, we  make use of the mode behavior in the taper region of an optical nanofiber to achieve 
a dichotomous regime where nanoparticles of one size are trapped while those of a different size are transported. This scheme constitutes a type of optical sieve, but unlike a conventional sieve, either larger or smaller particles can be retained while the other species passes through. Our method is significantly different to other recent results which demonstrated size selective optical forces~\cite{Ploschner, nan2018sorting, Shi} in that it achieves a dichotomous transport regime with a passive configuration, by making use of the mode behavior in the nanofiber taper.

\section{Physical principles}
\begin{figure}
\includegraphics[width=\linewidth]{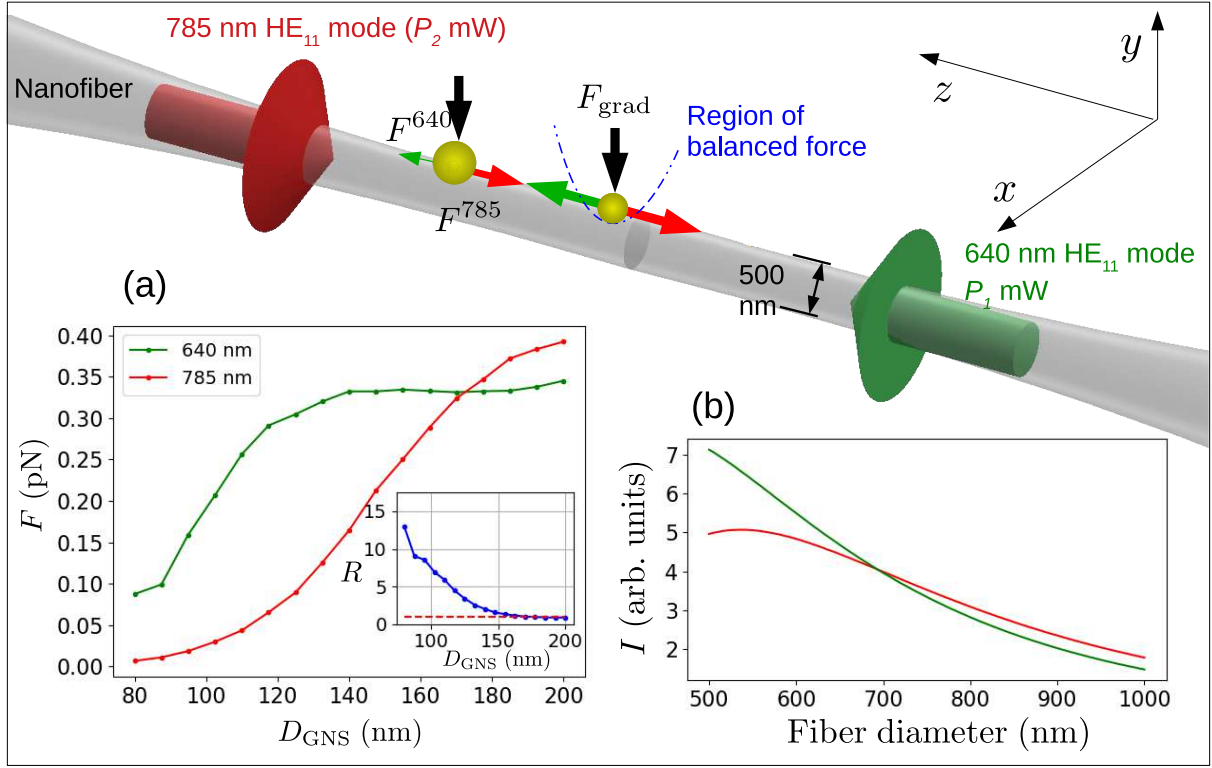}
\caption{\label{fig:principle}Concept and physical principles. Conceptual diagram showing the trapping and transport
of 100 nm and 150 nm diameter gold nanospheres along a 500 nm diameter optical nanofiber. Green and red arrows depict the optical force
due to the 640 nm HE$_{11}$ mode and the 785 nm HE$_{11}$ mode respectively. Inset (a) shows FDTD simulated force along the fiber axis for 640 nm light and 785 nm light as indicated in the legend, with the inset showing their ratio. Inset (b) shows
how the light intensity at the fiber surface depends on the fiber diameter.}
\end{figure}
The concept of our experiment is illustrated in Fig.~\ref{fig:principle}. Colloidal gold nanospheres (GNSs) which come close to an optical nanofiber are trapped near its surface by the gradient force $F_{\rm grad}$ and propelled by the absorption and scattering of photons from the evanescent portion of the fiber mode~\cite{kawata1996optically}. This force is $F^{640}$ in the positive $z$ direction due to the $+z$ propagating $x-$ polarized 640 nm HE$_{11}$ (fundamental)  mode, and $F^{785}$ in the negative $z$ direction due to the $-z$ propagating $x-$ polarized 785 nm HE$_{11}$ mode. These forces will balance at a certain power difference between the two beams which depends on the polarizability of the particles at both wavelengths. 

Inset (a) of Fig.~\ref{fig:principle} shows the forces experienced by a particle of diameter $D_{\rm GNS}$ on the
surface of a 550 nm diameter nanofiber for an input power of 1 mW at 640 nm wavelength (green points) and 785 nm wavelength (red points). 
In both cases, the data points show forces evaluated using finite difference time domain (FDTD) simulations of the system for 
taper parameters matched to experimental values (see Fig.~\ref{fig:exp}(a)), with lines connecting the points to guide the eye. The origin of size selectivity is apparent if we compare the ratio of forces $R=F^{640} / F^{785}$ at each wavelength as a function of particle size, as shown in the inset. For example, comparing forces on 100 nm and 150 nm diameter particles, we see that $R\approx7.5$ and $R\approx1.5$ respectively. This means that if we hold the power of the 785 nm mode constant, force balance will occur at a 640 nm mode power which is about five times lower for a 100 nm GNS relative to a 150 nm GNS.

Above the power for force balance, trapping occurs according to the mechanism shown in inset (b) of Fig.~\ref{fig:principle} which shows the calculated dependence of the electric field intensity at the nanofiber surface as a function of the fiber diameter for a 640 nm mode (green line) and 785 nm mode (red line). It may be seen that the two curves do not have the same gradient and cross over at a certain fiber diameter. This means that it is possible for the 640 nm  mode to exert the dominant force on a particle at one fiber diameter and the 785 nm mode to exert the (oppositely directed) dominant force at a larger diameter. For tapered fibers, this behavior either leads to an unstable potential maximum or a stable (trapping) potential minimum at some point along the taper, with the trap position moving to higher diameter as the power of the 640 nm beam is increased. We refer to this trap configuration as a \textit{two color taper trap} and note that it is distinct from two color trapping at constant nanofiber radius which has been extensively studied both theoretically and experimentally for atomic systems~\cite{Fam1} and theoretically in the case of nanoparticles~\cite{Xiao,skelton2012evanescent}. We will calculate the potential for an explicit example below.

\section{Numerical calculations}
\begin{figure}
\includegraphics[width=\linewidth]{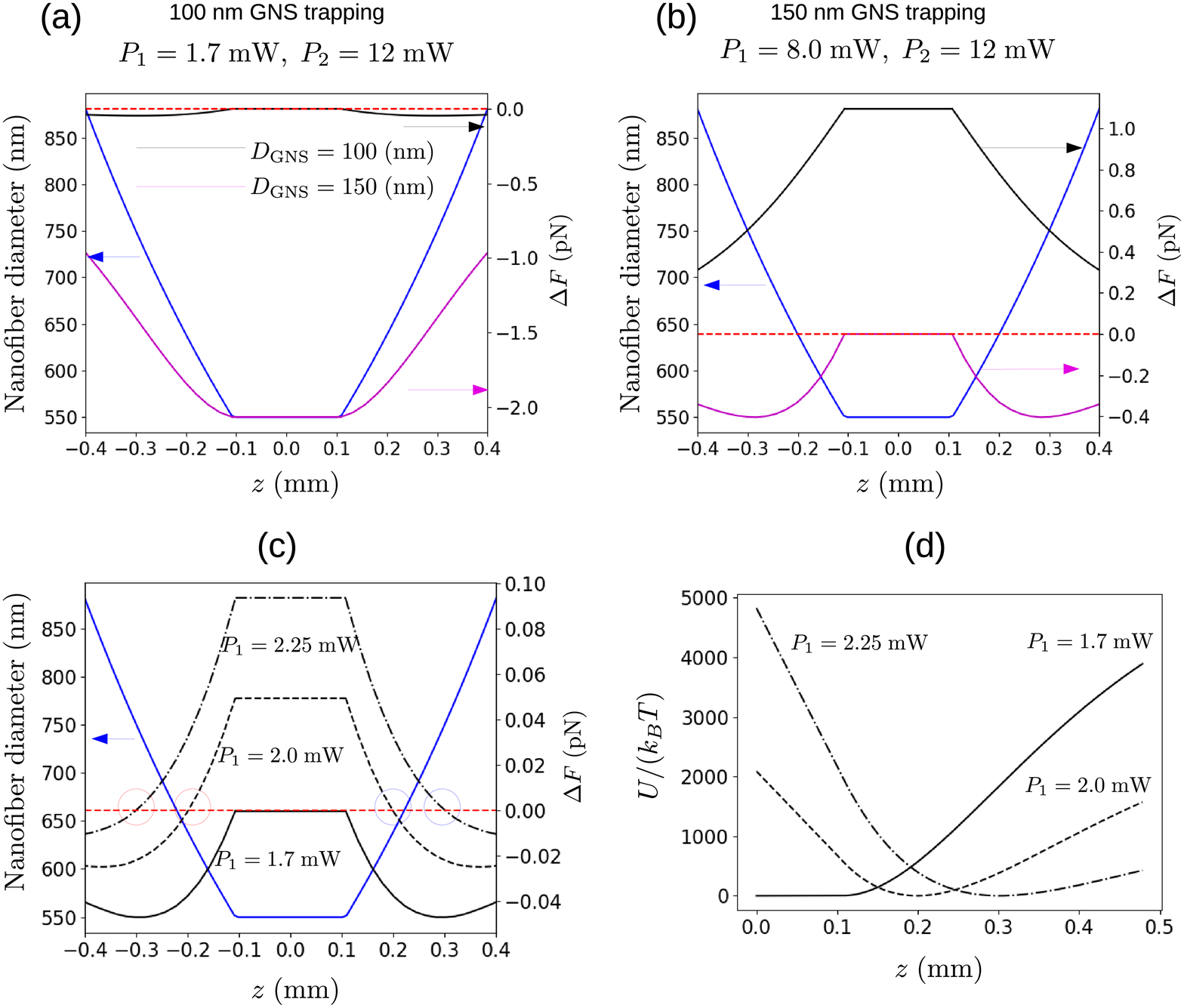}
\caption{\label{fig:trappingtheory}Numerical predictions. (a) and (b) show the dependence of optical forces along the nanofiber axis with $P_2=12$ mW (as in experiments) and given force balance conditions for a 100 nm diameter GNS and a 150 nm diameter GNS respectively. In all cases the blue curve shows the nanofiber taper diameter profile (left hand vertical axis) and black and magenta curves show optical forces on 100 nm and 150 nm diameter GNSs respectively (right hand vertical axis). (c) Variation of the force difference (i.e. total force) on a 100 nm GNS for $P_1 = 1.7$ mW (force balance condition, solid line), $P_1=2.0$ mW (dashed line) and $P_1=2.25$ mW (dash-dotted line). As before, the blue curve shows the nanofiber taper diameter profile (left hand vertical axis). Blue circles show points where stable trapping is possible, whereas red circles show points where anti-trapping (i.e. a local potential maximum) exists. (d) Shows the calculated potential energy of a 100 nm GNS as a function of $z$ on the positive side of the taper associated with $P_1 = 1.7$ mW (no trap, solid line), $P_1=2.0$ mW (dashed line), and $P_1=2.25$ mW (dash-dotted line). Note that the potential minimum has been assigned to zero energy in each case to allow simple comparison of the trap depths.}
\end{figure}
In order to experimentally demonstrate size selective trapping, we chose particles with a size difference of
50 nm - specifically 100 nm and 150 nm diameter GNSs. This relatively large size difference means that, assuming constant $P_2$, trapping occurs for very different $P_1$ for 100 nm as opposed to 150 nm diameter GNSs. This large power difference 
allows us to unambiguously identify trapping of the two different particle species, despite the fact that they are below the diffraction limit of our microscope. For smaller particle size differences, it would be necessary to use fluorescently tagged particles with different fluorescence wavelengths in order to confirm size selection, leading to a more complicated experimental setup.

We first performed FDTD simulations for the two different particle diameters to determine the likely trapping condition. Figures~\ref{fig:trappingtheory} (a) and (b) show FDTD calculated optical forces $\Delta F=F^{640} - F^{785}$ for the 
case of a 100 nm GNS (black line) and a 150 nm GNS (magenta line), with force balance conditions for 100 nm and 150 nm diameter GNSs respectively. The nanofiber diameter is set to 500 nm. With $P_2$ held at 12 mW, force balance is seen to occur at 1.7 mW in the 640 nm mode for 100 nm diameter GNSs and 8.0 mW in the 640 nm mode for 150 nm  diameter GNSs. 
Figure~\ref{fig:trappingtheory} (c) shows the force $\Delta F$ on a 100 nm diameter GNS as the power $P_1$ is raised beyond the force balance condition. It may be seen that the zero crossings of $\Delta F$ occur at certain positions on the negative and positive $z$ halves of the taper. On the $z<0$ side these zero crossings have an anti-trapping character, where movement away from the zero-force position leads to a growing force in the direction of movement. On the $z>0$ side of the taper, the zero-crossings have a trapping character where movement away from the zero position results in a restoring force which pushes the particle back towards the zero position. By numerically integrating the relation $\Delta F = -\partial U / \partial z$,  where $U$ is the potential energy of the particle, we can estimate the potential near the zero points for each power in Fig.~\ref{fig:trappingtheory}(c).
Figure~\ref{fig:trappingtheory}(d) shows the results of these calculations. For the case of force balance, $P_1=1.7$ mW (solid line), an inflection point rather than a potential minimum occurs. For $P=2.0$ mW and $2.25$ mW, a potential minimum is found at the zero force position, and is seen to move in the positive $z$ direction as the power is increased. 

\section{Experiment}
\begin{figure}
\includegraphics[width=\linewidth]{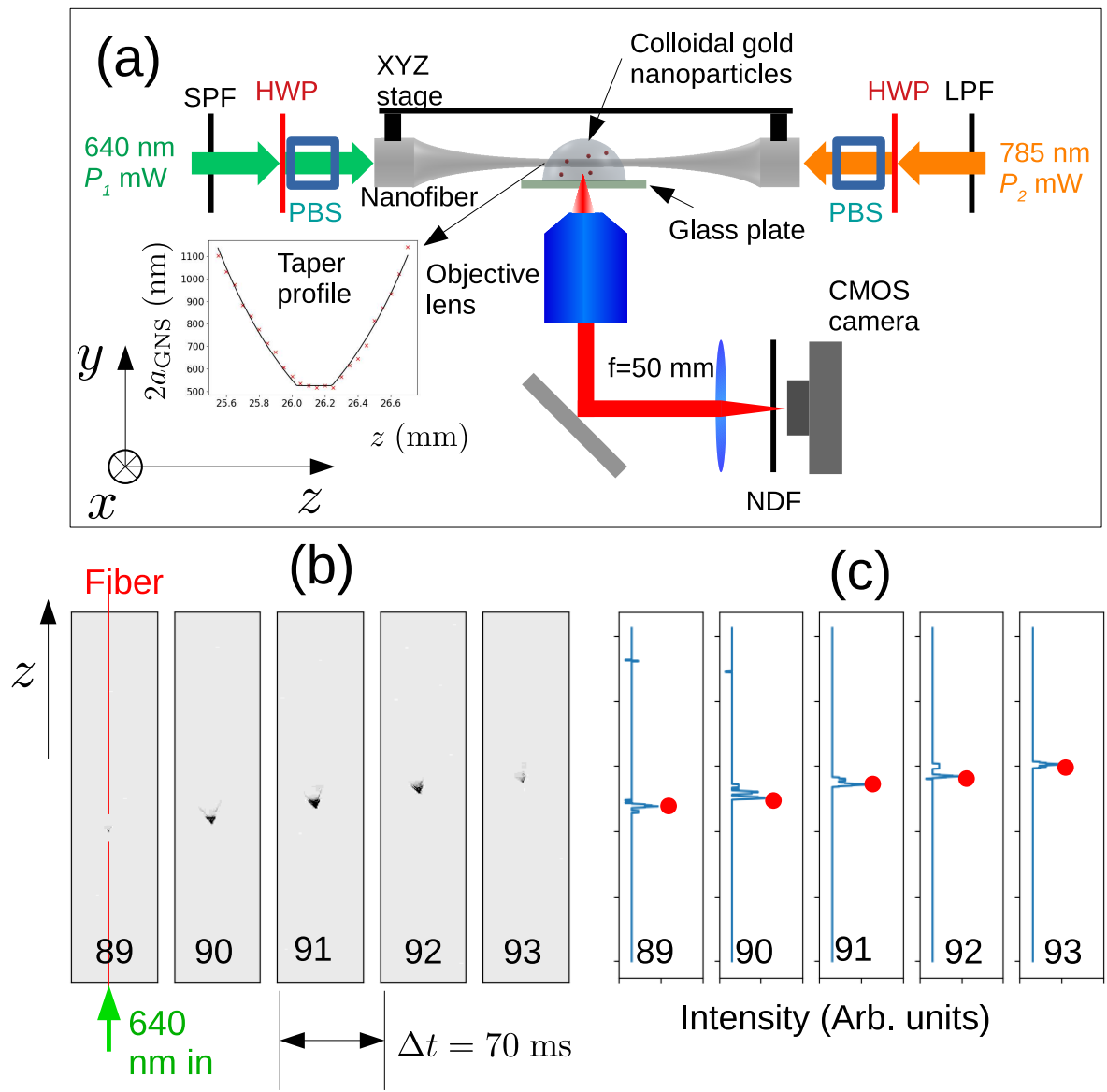}
\caption{\label{fig:exp}Experiment. (a) The experimental setup. An optical nanofiber was immersed in a droplet containing colloidal gold nanoparticles.
Counterpropagating 640 nm and 785 nm wavelength guided modes were introduced into the fiber from diode laser sources, with a short pass filter (SPF) and long pass filter (LPF) employed to keep counterpropagating light from entering the laser diodes. Each field passes through a half-wave plate (HWP) and a polarizing beamsplitter (PBS) before entering the fiber. Light scattered from particles trapped by the nanofiber guided modes was collected by an objective lens, attenuated by a neutral density filter (NDF) and detected by a complementary metal-oxide-semiconductor (CMOS) camera. The inset shows the scanning electron microscope measured nanofiber diameter. (b) Example frames from a particle transported along the nanofiber. The frame number is indicated at the bottom of each frame. (c) 1D data extracted from the frames in (b) by taking only the pictures which lie along the fiber line. Red dots indicate peaks, which are detected and used to create a reduced data set. (See text for details).}
\end{figure}
We now move on to the experiment. The experimental setup is as shown in Fig.~\ref{fig:exp}(a). An optical nanofiber, with a waist diameter of 550 nm, and waist length of 200 $\mu$m was immersed in a droplet of ultra-pure water. Colloidal gold nanoparticles (150 nm diameter: Nanopartz A11-150-CIT-DIH; 100 nm diameter: Nanopartz A11-100-CIT-DIH) were then added to this droplet as necessary. Counterpropagating 640 nm and 785 nm wavelength guided modes were introduced into the fiber from free running diode laser sources, with a short pass filter (SPF) and long pass filter (LPF) employed to keep counterpropagating light from entering the laser diodes. We polarized both fields before they entered the fiber by passing them through a half-wave plate and a polarizing beamsplitter, after which the fields were $x-$polarized, with an intensity controlled by the half-wave plate angle. Light scattered from particles trapped by the nanofiber guided modes was collected by a 20x objective lens (Olympus LU Plan Fluor, numerical aperture 0.45), attenuated by a neutral density filter (NDF), focused and then detected by a complementary metal-oxide-semiconductor (CMOS) camera (Thorlabs DCC1545M). The CMOS camera-measured intensities $I$ at each pixel constitute our raw data. Note that due to complex scattering of the guided mode light from the nanoparticles and the fiber itself, transported nanoparticles do not have a perfectly circular shape when imaged, even for optimal alignment of the objective lens. Nonetheless, particles can be clearly identified in the data. Example frames are shown in Fig.~\ref{fig:exp}(b), for the situation where $P_1=12$ mW and $P_2=0$ (i.e. no 785 nm mode). A trapped, transported particle is seen as a scattering point which moves along the fiber in the same direction as the 640 nm mode. Because the motion is effectively 1D, we can simplify our data considerably by only considering pixels along the line corresponding to the fiber position. Fig.~\ref{fig:exp}(c) shows such 1D data for each frame of Fig.~\ref{fig:exp}(b). 

Our experimental strategy to observe size-selective trapping using the apparatus described above, is to first observe the onset of trapping for a solution containing \emph{only} 150 nm diameter GNSs. Once the power dependence of transport behavior for 150 nm diameter GNSs has been characterized, we add 100 nm diameter colloidal GNSs to the solution and repeat the experiment. 

\begin{figure*}
\includegraphics[width=\linewidth]{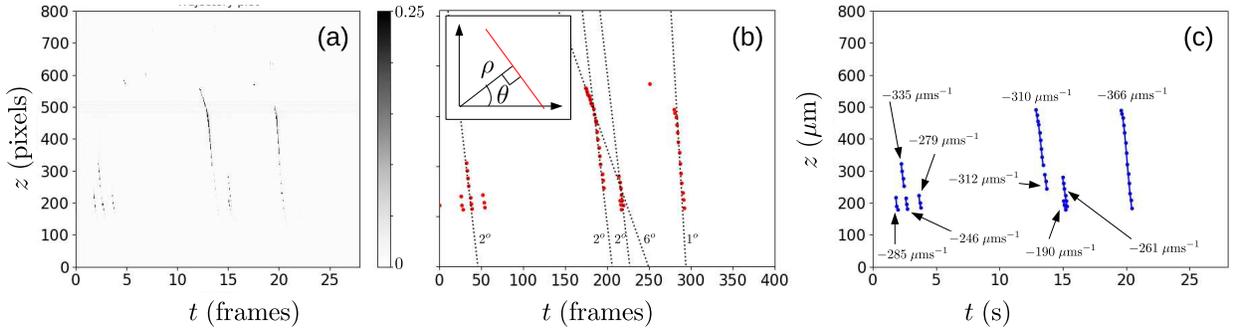}
\caption{\label{fig:result1} Experimental data: transport of 150 nm diameter GNSs. (a) Combined 1D cross-sections for each captured frame in the data set for parameters $P_1=0$ mW, $P_2=12$ mW, and 150 nm GNSs only. (b) Detected peaks (red points) in 1D cross section data along with 
five lines detected by a Hough analysis of the data (dashed black lines). The inset depicts the polar representation of a line (shown in red) used
by the Hough analysis, and the detected angle $\theta$ is shown beside each detected line. (c) GNS trajectories detected in the data using the correlation technique explained in the text. The associated mean velocity is shown beside each trajectory. }
\end{figure*}
We now move on to our experimental results. First, we consider the case of purely transporting motion for a solution of 150 nm GNSs in the presence of 12 mW of the 785 nm HE$_{11}$ mode alone. After introducing the fiber to a pure water droplet we added 150 nm GNS producing a concentration of 2.6$\times10^5$ particles / $\mu$L. We then took 400 frames of data as an avi file using the CMOS camera. 
The frames were extracted to individual image files and a 1D cross-section of each frame was taken along the line corresponding to the fiber position. Combining this 1D data into a single intensity matrix $I(f,p)$ (where $f$ is an integer specifying the frame and $p$ is an integer specifying the pixel) gives Fig.~\ref{fig:result1}(a), where the horizontal axis is the time in frames, and the vertical axis is the distance along the fiber axis in pixels. 

We then used a peak extraction algorithm (Python Scipy module \textit{find\_peaks}) to find the estimated position of the particle in each frame, resulting in a reduced data set $(l,m)$ of frame $l$ and pixel position $m$ for each detected peak. These positions are shown by red dots in Fig.~\ref{fig:result1}(b). From the peak data, we used two methods to analyze the particle trajectories along the fiber. The first was the Hough-transform method~\cite{hough}.  
Using a polar parameterization of a straight line $(\rho,\theta)$, as shown in the inset of Fig.~\ref{fig:result1}(b), the Hough transform creates a spectrum of the peak data in $\rho-\theta$ space. The peaks of the spectrum correspond to lines connecting points in the data, with the largest peak corresponding to the line containing the most points. In this sense, the Hough transform objectively identifies the dominant trajectory in the data. Sample lines found by the Hough transform are shown as dashed black lines in Fig.~\ref{fig:result1}(b), and their $\theta$ values are indicated beside each line. 

The second method we used was a simple correlation method to find nearest peaks in consecutive frames giving the trajectories shown in Fig.~\ref{fig:result1}(c), where the axes have been converted to time in s and distance in $\mu$m by application of the relevant calibration factors. The calculated mean velocity is shown beside each trajectory. We used this method to find the mean and standard deviation of trajectories in each data set.

The use of the above data analysis methods allows us to perform an objective analysis of our data rather than focusing on individual trajectories which are subjectively judged to be of interest.

\begin{figure*}
\includegraphics[width=\linewidth]{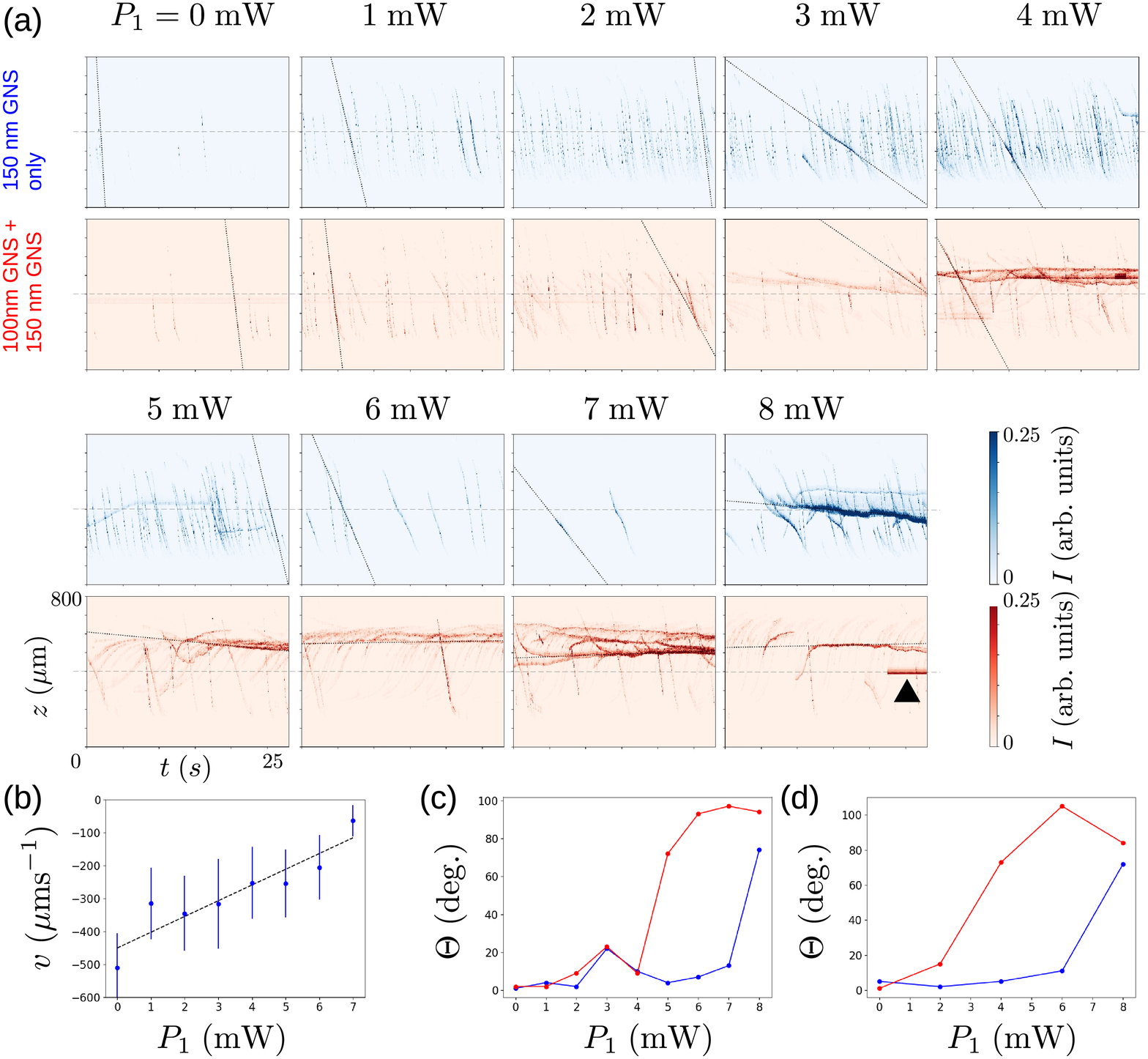}
\caption{\label{fig:result2}Experimental data: size selective trapping. (a) (Upper rows) Power dependence of transport for 150 nm diameter GNSs,
with $P_1$ as indicated and $P_2$ fixed at 12 mW. The onset of trapping is observed at $P_1=8$ mW. (Lower rows) Power dependence of transport for a mixture of 100 nm diameter and 150 nm diameter GNSs. The onset of trapping is now observed at $P_1=3$ mW, corresponding to trapping conditions for the 100 nm diameter GNSs. In each case, a black dotted line shows the dominant line of the Hough spectrum, while dashed horizontal lines indicate the nominal position of the nanofiber center. The black arrow in the lower row of the 8 mW data indicates a particle stuck to the fiber surface. This was ignored for the purposes of the Hough analysis.  (\textbf{See movie file in Supplementary Material}) (b) Power dependence of the mean velocity for trajectories detected in each data set shown in the upper rows of (a). Error bars show $\pm$ one standard deviation. (c) Angle $\Theta$ of the top ranked line detected by Hough analysis for each data set shown in the upper rows of (a) (blue points) and the bottom row of (a) (red points). (d) Same as (c), but for a data set where there was no control over the polarization of the fields entering the nanofiber.}
\end{figure*}
We first performed measurements in the presence of 150 nm diameter GNSs only, with the same concentration as given above. The power in the 785 nm mode was kept constant at $P_2=12$ mW, while the power of the 640 nm mode was varied from 0 to 8 mW in 1 mW increments. Immediately after setting the power, 400 frames of data was taken. Results are shown in the upper rows of 
Fig.~\ref{fig:result2}(a). Note that all plots in Fig.~\ref{fig:result2}(a) have the same axes as shown for the $P_1=5$ mW data in the lower row.
Only transporting trajectories in the direction of the 785 nm mode propagation were seen until $P_1$ reached 8 mW. At this point, trapped particles were observed as can be clearly seen by the near horizontal lines in the combined data $I(f,p)$.

After taking the data shown in the upper row of Fig.~\ref{fig:result2}(a), we added 100 nm diameter GNSs solution to the droplet giving a concentration of $2.2\times 10^5$ particles / $\mu$L for the already present 150 nm GNSs and $6.3\times 10^5$ particles / $\mu$L for the 100 nm GNSs. No change in the transmitted power of either wavelength mode was observed due to the addition of the solution. 
We then performed the experiment again at each value of $P_1$. This time, the onset of particle trapping occurred at approximately 3 mW, as judged by the appearance of multiple near horizontal particle trajectories (lower row, Fig.~\ref{fig:result2}(a)). Inspection of the data in the lower row of Fig.~\ref{fig:result2}(a) shows interesting behavior beginning at $P_1=5$ mW, where particle trajectories travelling in both directions along the fiber can be seen. For $P_1=8$ mW, the data suggests that many particles travel through the region where other particles are trapped.  This is due to 100 nm diameter GNSs being transported along the fiber due to the stronger force they experience from the 640 nm mode. At the same value of $P_1$ 150 nm GNSs are trapped, and thus the behavior constitutes a type of optical sieve effect, but one where either smaller or larger particles can be retained depending on the adjustable optical parameters.  (This behavior is easier to perceive by looking at the movie file provided in the supplementary material).

We note that in the lower row of the 8 mW data, a particle adhering to the fiber is observed. Irreversible particle sticking happened at a rate of approximately once in two hours, and is distinguishable from particle trapping due to the constant position and constant intensity of the scattering observed.

We also note that although a faint horizontal trajectory can be see in the 150 nm diameter GNS data at 5 mW, this does not mark the onset of trapping. The reason is that trapping is not seen in the following 6 mW or 7 mW data. Because trapping always exists for a range of powers above the onset power, this rules out 5 mW as the onset of trapping for 150 nm GNSs. Instead the trajectory seen in the 5 mW case is most likely due to a rare smaller diameter particle in the solution.

Fig.~\ref{fig:result2}(b) shows the results of a trajectory analysis of the 150 nm only data, which reveals how the mean trajectory velocity reduced to near zero as $P_1$ was increased. Error bars show plus and minus one standard deviation of the velocity over the observed trajectories. The standard deviation may be seen to be rather large. Variation in the observed velocities is at least partly 
attributed to the variation in particle size which is specified to be $\pm 10$ nm.

Fig.~\ref{fig:result2}(c) shows the angle $\Theta$ at the peak of the Hough spectrum for each data set shown in Fig.~\ref{fig:result2}(a). Note that 
the dominant line in the Hough spectrum for each data set is shown by a black dotted line in Fig.~\ref{fig:result2}(a).
The tendency of this value to $90^o$ with the onset of trapping is a quantitative measure of the fact that in the trapping regime, 
particle trajectories are dominated by long lifetime trajectories with near-zero velocity. The behavior of $\Theta$ is clearly different for the case with only
150 nm diameter GNSs and the case where both 150 nm and 100 nm diameter GNSs are present. 
We also made similar measurements for a different
fiber without polarization control. The Hough angles for these results, shown in Fig.~\ref{fig:result2}(d), show very similar qualitative behavior, suggesting the repeatability of the size-selective trapping behavior. The results also show that polarization control is not necessary to achieve the size-selective trapping effect, which is convenient for practical applications.

\section{Discussion}

Let us compare our experimental results with the numerical predictions. First we consider qualitative predictions.
According to the principle of the two-color technique, the onset of trapping will occur near to the nanofiber center (i.e. $\pm 100\;\mu$m about the dashed line shown in Fig.~\ref{fig:result2}(a), where the nanofiber diameter is constant), as seen in the case of 150 nm diameter GNSs only when $P_1=8$ mW and also for $P_1=3$ mW when 100 nm diameter GNSs are present. As $P_1$ is increased, the position of trapping should move in the positive $z$  direction, and no trapping should be observed below the nanofiber center. Our experimental results are completely in line with this theoretical prediction. Most importantly, the onset of trapping for 100 nm GNSs should occur at a much lower power for 100 nm GNSs than for 150 nm GNSs.
Again, our data are consistent with this prediction.

In terms of quantitative predictions, the onset of trapping is seen to occur at $P_1=8$ mW for 150 nm diameter GNSs only, in agreement with the numerical prediction of 8.0 mW. When 100 nm GNSs are present, the numerics predicts an onset of trapping at 1.7 mW i.e. 2.0 mW at the 1 mW resolution of the experiment. The onset of trapping is seen at 3 mW in experiments - which may be due to a number of effects including the difference in plasmon resonance lineshape for the experimentally used 100 nm GNSs relative to the ideal simulated versions, along with non-optical effects such as heating (not included in our simulations).
The numerics also predicts that the taper trapping region should be between 100 and 300 $mu$m from the nanofiber center,
in line with what was observed experimentally in the case where 100 nm GNSs are present.

Although the evidence for size-dependent trapping and the optical sieve effect accumulated here is necessarily indirect, due to the sub-wavelength size of the nanoparticles, the above agreement between multiple aspects of theory, numerics and experiment provides strong support for our interpretation. We also note that using ab objective measure of the dominant trajectory in each data set (i.e. the Hough analysis), rather than relying on our visual impression of the data clearly shows the expected difference between the onset of trapping in the case where only 150 nm GNSs are present, compared to the case where both 100 nm and 150 nm diameter GNSs are in the solution.

In future experiments, we anticipate that similar size selective effects could be achieved for non-metallic particles, using, for example, whispering gallery resonances, or other size dependence resonance effects. Furthermore, our current demonstration was for a 50 nm difference in particle diameters, but simulations suggest that selective trapping for differences down to 10 nm is possible. The simplicity and robustness of the experimental setup used here makes it an excellent candidate for practical nanoparticle sorting applications.

\section*{Acknowledgement}
M.S. acknowledges support from JSPS KAKENHI (Grant no. JP19H04668) in Scientific Research on Inovative Areas ``Nano-material optical-manipulation".

\bibliography{SizeSelective}

\end{document}